\documentclass[aps,pra,onecolumn,balancelastpage,amsmath,amssymb,showpacs]{revtex4}
\usepackage{dsfont}
\usepackage{color}         
\usepackage{graphics}      
\usepackage[pdftex]{graphicx}      
\usepackage{longtable}     
\usepackage{epsf}          
\usepackage{epstopdf}
\DeclareGraphicsExtensions{.eps}
\usepackage{bm}            
\usepackage{thumbpdf}
\usepackage{hyperref}
\hypersetup{
	colorlinks=true,
	allcolors=blue}							  
\usepackage[utf8]{inputenc}
\makeatletter

\renewcommand*{\p@subsection}{}

\renewcommand*{\p@subsubsection}{}
\makeatother

\begin{document}
\title{A Study of Shell Model Neutron States in $^{207,209}Pb$ Using the Generalized
 Woods-Saxon plus Spin-Orbit Potential}
\author{J. A. Liendo}
\email{jliendo@usb.ve}
\author{E. Castro}
\author{R. Gómez}
\date{\today}
\affiliation{Physics Department, Sim\'{o}n Bol\'{i}var University, Apdo. 89000, Caracas 1086,
Venezuela}
\author{D. D. Caussyn}
\affiliation{Physics Department, Florida State University, Tallahassee, Florida 32306, USA}

\begin{abstract}
The experimental binding energies of single-particle and single-hole neutron states
belonging to neutron shells that extend from N = 126 to 184 and 82 to 126 respectively,
have been reproduced by solving the Schr\"{o}dinger  equation with a potential that has
two components: the generalized Woods-Saxon (GWS) potential and the  spin-orbit (SO)
coupling term. The GWS potential contains the traditional WS potential plus a term (SU) whose
intensity reaches a maximum  in the nuclear surface. Our results  indicate the existence
of a explicit relationship between the strength of the SU potential and the orbital
angular momentum quantum number  $\ell$  of the state. This dependence has been used to
make reasonable predictions for the excitation energy centroids of states located inside
and outside the neutron shells investigated. Comparisons are made with results reported
in previous investigations.
\end{abstract}

\maketitle

\section{Introduction}

Single-particle and single-hole neutron states have been previously investigated in the
region around the doubly-magic $^{208}Pb$ nucleus
 \cite{Blomqvist,Rost1,Dudek1,Dudek2,Volya,Wang}. The interaction of a neutron with the
rest of the nucleus referred to as the core has been represented in these studies by a
Hamiltonian containing a nuclear Woods-Saxon (WS) potential\cite{WS} and a spin-orbit
(SO) coupling term. Although the same potential parameterization has been used in
Refs.~\cite{Blomqvist,Rost1,Dudek1,Dudek2}, a unique set of parameter values has not been
found. In fact, none of the reported parameterizations, including the one published by
Schwierz et al.\cite{Volya}, has been able to produce an overall agreement between the
predicted single-particle and single-hole neutron state energies and the corresponding
experimental values.

In this work, we have used the so called generalized Woods-Saxon (GWS) potential instead
of the original WS potential\cite{WS}, with the expectation of reproducing the
experimental binding energies of single-particle and single-hole neutron orbitals that
exist in the neutron shells $N = 126 - 184$ and $82-126$, respectively. This potential
contains the WS potential plus a term referred to as the surface (SU) potential that
maximizes in the nuclear surface and is linearly proportional to the derivative of a WS
function.  It is a well known fact that the WS potential alone does not reproduce the
energies of $\ell$ = 0 single-particle levels with enough accuracy when applied to a wide
nuclidic region \cite{Gonul}. Using a formalism where the Schr\"{o}dinger  equation is
expressed in terms of the Jacobi polynomials, G\"{o}n\"ul et al.\cite{Gonul} have
obtained the total potential influencing a single $\ell$ = 0  neutron in a nucleus. The
deduced potential (a GWS potential) contains a WS potential plus an additional term (a SU
potential) that provides the flexibility to construct the surface structure of the
related nucleus. The validity of the theoretical treatment presented in Ref. \cite{Gonul}
is supported by the careful analysis of the analytical results obtained and the detailed
discussion of the formalism prescriptions followed.

The GWS potential has been the object of study in different works published in the
literature. Hamzavi et al. \cite{Hamzavi} have obtained solutions of the Dirac equation
for a nucleon experiencing the GWS potential under the relativistic spin symmetry limit.
Bayrak et al. \cite{Bayrak1}  have modelled the behavior of a neutral pion in a nucleus
and obtained an analytical solution of the Klein-Gordon equation for a spin = 0 particle
moving in the GWS potential field. Using the Dirac equation, Candemir et al.
\cite{Candemir} have determined negative energy eigenvalues which satisfy the boundary
condition for any $\kappa$ (spin-orbit operator eigenvalue) states of an anti-proton
being subjected to the GWS potential in the pseudospin symmetry limit. Also, they have
calculated the bound state energy eigenvalues of a proton moving under the GWS potential
in the spin symmetry limit. These authors point out the usefulness of the SU potential to
examine the single-particle energy levels of a nucleon and an anti-nucleon since the
interactions of the surface nucleon or anti-nucleon are very important in explaining the
energy spectrum of the nuclei.

The SU term of the GWS potential has been shown in some cases to generate an additional
potential pocket in the nuclear surface region that is crucial to understand the elastic
scattering of several nuclear reactions \cite{Brandan, Satchler}. In other cases, the SU
potential does not induce this potential pocket as in the heavy ion elastic scattering
analysis of I. Boztosun \cite{Boztosun} where the inclusion, within the framework of the
optical potential, of two small real SU potentials, was absolutely necessary to reproduce
the elastic scattering data especially at large angles. In this case, although the SU
potentials do not produce pockets, they generate interference effects that lead to an
impressive reproduction of the experimental cross section oscillations \cite{Boztosun}.
The potential pocket mentioned here has also been obtained theoretically by Koura et
al.~\cite{Koura} from the study of a refined WS potential that has a term that increases
the freedom in the surface structure of the potential.

C. Berkdemir et al.~\cite{Berkdemir1} used the Schr\"{o}dinger equation with the GWS
potential to generate the binding energies of several $\ell$ = 0 states of a hypothetical
nucleus formed by a valence neutron plus an inert core of 56 nucleons. Unfortunately, the
eigenvalue equation reported in this publication was later on shown to be physically
incorrect~\cite{Berkdemir2}. However, the numerical calculations included in
Ref.~\cite{Berkdemir2} show that the addition of the SU term to the WS potential allows
the modification of the binding energies of $\ell = 0$ states that otherwise would not be
possible with the addition of a SO potential only.  Recently, Bayrak et
al.~\cite{Bayrak2} examined the bound state solution of the radial Schr\"{o}dinger
equation with the GWS potential using the Pekeris approximation in terms of an
appropriate boundary condition for arbitrary  $\ell$ states. From the closed form
obtained for the energy eigenvalues, these authors calculated single-particle energies
corresponding to a neutron orbiting around the $^{56}Fe$ nucleus that are comparable to
the numerical results for $\ell$ = 0.

In this work, we use the GWS+SO potential mentioned above (GWS = WS + SU) to reproduce
the experimental binding energies of zero and nonzero $\ell$ states in $^{209}Pb$ and
$^{207}Pb$. The WS and SO parameter values have been taken from the work published by
Schwierz el al.~\cite{Volya}. For every state analyzed, we have obtained the strength of
the SU potential that makes coincide the theoretical binding energy with the
corresponding measured value. The experimental binding energies have been extracted from
previously reported neutron transfer reaction experiments carried out on
$^{208}Pb$~\cite{Martin,Kovar,Gales,Matoba} and the theoretical binding energies have
been obtained by solving the Schr\"{o}dinger equation with the computer program
GAMOW~\cite{Gamow}. The strength of the SU potential has been shown to be a function of
the orbital angular momentum of the state. This dependence has been used to predict the
excitation energies of some states that lie inside and outside the neutron shells
investigated.

\section{The Traditional Woods-Saxon plus Spin-Orbit Potential}
\subsection{Formalism}
As a starting point, we used a nuclear potential given by
\begin{equation}
V(r)= V_{WS}(r) + V_{SO}(r),
\end{equation}
where the $V_{WS}(r)$ and  $V_{SO}(r)$ potentials have the analytical expressions
\begin{equation}
V_{WS}(r)= - \frac{V_{WS}}{1 + \exp \left(\frac{r - R}{a}\right)},
\end{equation}
and
\begin{equation}
V_{SO}(r)=  - \frac{\lambda}{4} {\lambda_{C}}^{2} \frac{1}{r} \frac{d\tilde{V}_{SO}}{dr}
\left[j(j+1)-\ell(\ell+1)-\frac{3}{4}\right].
\end{equation}
Here,  $\lambda$ is a dimensionless parameter related to the depth of the SO potential
and $\lambda_{C}$ is the reduced Compton wavelength of the nucleon-core system given in
fm units by

\begin{table}[t]
\caption{\label{tab:STAB1}Experimental (${^{s.p.}B}_{n \ell j}$) and theoretical
(${^{s.p.}B}_{th, n \ell j}$) binding energies of single-particle neutron states in
$^{209}Pb$.  The experimental binding energies were obtained from the expression
${^{s.p.}B}_{n \ell j}= -3.9374 + E_{n \ell j}(209)$, where the energy centroid of each
orbital, $E_{n \ell j}(209)$, is assumed to be equal to the excitation energy of only one
excited state in $^{209}Pb$ since the spectroscopic factor ( $C^{2}S/(2j + 1)$) measured
by Kovar et al.~\cite{Kovar} for each $^{209}Pb$ state, is close to 1.}
{\begin{tabular}{@{}llllll}
        \toprule
    orbital& $\ell$ & $E_{n \ell j}$(209)& $C^{2}S/(2j+1)$ & ${^{s.p.}B}_{n \ell j}$& ${^{s.p.}B}_{th, n \ell j}$\\
        (n$\ell$j) &    &(MeV)& &(MeV)  &(MeV)\\
        \colrule
    $2g_{9/2}$  & 4 & 0.000 & 0.83 & -3.937 & -3.822 \\
        $1i_{11/2}$ & 6 & 0.779 & 0.86 & -3.158 & -2.699 \\
        $3d_{5/2}$  & 2 & 1.565 & 0.98 & -2.372 & -1.915 \\
        $4s_{1/2}$  & 0 & 2.033 & 0.98 & -1.904 & -1.382 \\
        $2g_{7/2}$  & 4 & 2.492 & 1.05 & -1.445 & -1.117 \\
        $3d_{3/2}$  & 2 & 2.537 & 1.07 & -1.400 & -0.945 \\
        \botrule
\end{tabular}}
\end{table}

\begin{eqnarray}
\lambda_{C}=\frac{\hbar}{\mu c} = 0.210019\left( 1 + \frac{m_{n}}{M_{core}}\right)\ \ ,
\\
\tilde{V}_{SO}(r)=-\frac{V_{SO}}{1+\exp\left(\frac{r-R_{SO}}{a_{SO}}\right)}\ \ ,
\\
R = r_{0}  A^{1/3}\ \ , \ \ \mbox{ and \ \ }  R_{SO} = r_{SO}  A^{1/3}\ \ .
\end{eqnarray}
In these equations, $\mu$ represents the reduced mass of the neutron-core system, $m_{n}$
and $M_{core}$ are the neutron and core masses respectively, $\hbar$ is Planck's constant
divided by 2$\pi$, and c is the speed of light. The quantum numbers $\ell$ and $j$
correspond to the orbital and total angular momenta respectively of a single-particle or
a single-hole state. $V_{WS}$ ,  $R$ and  $a$ represent the depth, width and surface
diffuseness respectively of the WS potential. The parameters corresponding to the SO
potential are $\lambda$ $V_{SO}$ , $R_{SO}$, and $a_{SO}$ respectively. We selected the
parameterization published  by Schwierz el al. \cite{Volya}

\begin{equation}
V_{WS} = V_{0}(1 + \frac{\kappa}{A}[-(N - Z + 1 ) + 2])  \ \ , \ \  V_{SO} = V_{0} \ \ ,
\end{equation}
and the parameter values: $V_{0}= 52.06$ MeV, $\kappa = 0.639$, $r_{0} = 1.260$ fm,
$r_{SO} = 1.16$ fm, $a = a_{SO}= 0.662$ fm and $\lambda = 24.1$. This WS + SO
parameterization is applicable over the whole nuclear chart for nuclides between $^{16}O$
and the heaviest elements, and provides a relatively good description of the nuclear mean
field leading to quality single-particle and single-hole spectra, nuclear radii,
prediction of drip-lines, shell closures and other properties.

To calculate the binding energy of single-particle neutron states, the $^{209}Pb$ nucleus
($A=209$, $Z=82$ and $N=127$) is considered as a system formed by a valence neutron and a
core of 208 nucleons, equivalent to the doubly-magic $^{208}Pb$ nucleus. The reduced
Compton wavelength (Eq. (4)) is calculated with $m_{n}$ = 1.0087 u  and $M_{core}$ =
207.9767 u. For the determination of the single-hole neutron state energies, the
$^{207}Pb$ nucleus is treated as a $^{208}Pb$ nucleus minus one neutron or, equivalently,
a core of 207 nucleons ($M_{core}$ = 206.9759 u) plus a neutron-hole. Since a single-hole
state can be considered as an unoccupied single-particle state, the WS and SO widths and
strengths, used to obtain the binding energies of hole (unoccupied) states in $^{207}Pb$,
are determined with the values $A=208$, $Z=82$ and $N=126$.

\begin{table}[t]
  \caption{\label{tab:STAB2}A comparison between the experimental (${^{s.p.}B}_{n \ell j}$) and theoretical (${^{s.p.}B}_{th,n \ell j}$) binding energies of the $1j_{15/2}$ orbital in $^{209}Pb$  determined for different fragmentation schemes. The experimental energies were determined by evaluating Eq. (8).  The theoretical energies were calculated with  the computer program GAMOW \cite{Gamow} using the potential given in Eq. (1). The excitation energies of the  $^{209}Pb$ fragments, $E_{x}$(209), and their spectroscopic factors,  $C^{2}$ S / (2j + 1), were extracted from an investigation of the $^{208}Pb(d,p)$ reaction at $E_{lab}(d) = 20$ MeV  \cite{Kovar}. For the scheme 1, the energy centroid $E_{n \ell j}$(209) was assumed to be  equal to the $^{209}Pb$ excitation energy $E_{x}$(209)= 1.424 MeV. The schemes 2 and 3 include four $^{209}Pb$ fragments. The deuteron break up potential was used to generate the results corresponding to the scheme 3.}
{\begin{tabular}{@{}*{6}{l}}
        \toprule
    Fragmentation & $E_{x}$(209)& $C^{2} S / (2j + 1)$ & $E_{n \ell j}$(209)& ${^{s.p.}B}_{n \ell j}$& ${^{s.p.}B}_{th, n \ell j}$\\
        Scheme & (MeV)& &(MeV)&(MeV)&(MeV)\\
        \colrule
        1  & 1.424 & 1.00  & 1.424 & -2.513 & -2.369\\
        2  & 1.424 & 0.58 &1.782 & -2.155 & -2.369  \\
           & 3.052 & 0.070 & & & \\
           & 3.556 & 0.032 & & & \\
           & 3.716 & 0.032 & & & \\
        3  & 1.424 & 0.77  &1.769 &-2.168 & -2.369 \\
           & 3.052 & 0.09  &  & & \\
           & 3.556 & 0.04  & & & \\
           & 3.716 & 0.04  & & & \\
         \botrule
        \end{tabular}}
       \end{table}

\begin{table}[t]
\caption{ \label{tab:STAB3}Experimental (${^{s.h.}B}_{n \ell j}$) and theoretical
(${^{s.h.}B}_{th, n \ell j}$) binding energies of single-hole neutron states in
$^{207}Pb$. The experimental binding energies were obtained from the expression
${^{s.h.}B}_{n \ell j}$ = -7.3678 - $E_{n \ell j}$(207), where $E_{n \ell j}$(207)
represents the excitation energy centroid of each orbital. The  excitation energies of
the $^{207}Pb$ fragments, $E_{x}$(207), and their spectroscopic factors, $C^{2}S/(2j+1)$,
were extracted from Ref. \cite{Matoba} for the  $3p_{1/2}$, $2f_{5/2}$  and $3p_{3/2}$
states and from  Ref. \cite{Gales} for the rest of the states.}
{\begin{tabular}{@{}*{7}{l}}
 \toprule
orbital& $\ell$ & $E_{x}$(207)& $C^{2} S / (2j + 1) $& $E_{n \ell j}$(207)& ${^{s.h.}B}_{n \ell j}$& ${^{s.h.}B}_{th, n \ell j}$\\
     (n$\ell$j) & & (MeV)& &(MeV)&(MeV)&(MeV)\\
 \colrule
$3p_{1/2}$ & 1 & 0.000  & 1.1   & 0.000 & -7.368 & -7.687\\

$2f_{5/2}$ & 3 & 0.570  & 0.88  & 0.570 & -7.938 & -8.412\\

$3p_{3/2}$ & 1 & 0.900  & 0.96  & 0.900 & -8.268 & -8.492\\

$1i_{13/2}$ & 6 & 1.629 & 0.857  & 1.869 & -9.237 & -9.332 \\
            &   & 5.990 & 0.050  &       &        &        \\
$2f_{7/2}$  & 3 & 2.334  & 0.913 & 2.913 & -10.281& -10.716 \\
            &   & 4.572  & 0.044 &       &        &         \\
            &   & 4.765  & 0.035 &       &        &        \\
            &   & 6.370  & 0.113 &       &        &        \\
$1h_{9/2}$  & 5 & 3.415  & 0.690 & 4.014 & -11.382& -10.681  \\
            &   & 3.660  & 0.090 &       &        &       \\
            &   & 5.410  & 0.112 &       &        &        \\
            &   & 5.620  & 0.180 &       &        &        \\
$1h_{11/2}$ & 5 & 7.010  & 0.088 & 8.077 & -15.445& -15.906 \\
            &   & 7.590  & 0.088 &       &        &        \\
            &   & 7.960  & 0.064 &       &        &        \\
            &   & 8.220  & 0.055 &       &        &        \\
            &   & 8.540  & 0.077 &       &        &        \\
            &   & 9.220  & 0.088 &       &        &        \\
 \botrule
\end{tabular}}

\end{table}

The single-particle neutron orbitals studied in this work ($2g_{9/2}$, $1i_{11/2}$,
$1j_{15/2}$, $3d_{5/2}$, $4s_{1/2}$, $2g_{7/2}$ and $3d_{3/2}$) lie above the Fermi
level, belong to a neutron shell that extends from $N = 126$ to $182$, and all of them
except the $1j_{15/2}$ orbital have even $\ell$ values and positive parity. In contrast,
the single-hole neutron orbitals ($1h_{11/2}$,$1h_{9/2}$, $2f_{7/2}$, $1i_{13/2}$,
$3p_{3/2}$, $2f_{5/2}$ and $3p_{1/2}$) are located below the Fermi level, all of them
except the $1h_{11/2}$ level belong to the $N=82-126$ shell, and all of them except the
$1i_{13/2}$ orbital  have odd $\ell$ values and negative parity. We used the computer
code GAMOW~\cite{Gamow} with some minor modifications to calculate the binding energies
of all these orbitals. The theoretical results obtained have been compared with the
experimental single-particle and single-hole binding energies, ${^{s.p.}B}_{n \ell j}$
and ${^{s.h.}B}_{n \ell j}$, given respectively by

\begin{equation}
{^{s.p.}B}_{n \ell j}= \Delta M(209) - \Delta M(208) - \Delta M_{n} +  E_{n \ell j}(209)
\ \ ,
\end{equation}
and
\begin{equation}
{^{s.h.}B}_{n\ell j}=  \Delta M(208) - \Delta M(207) - \Delta M_{n} - E_{n \ell j} (207)
\ \ ,
\end{equation}
where  $\Delta$$M_{n}$,   $\Delta$M(209),   $\Delta$M(208)  and  $\Delta$M(207) are the
neutron, $^{209}Pb$,  $^{208}Pb$ and  $^{207}Pb$  mass excesses equal to 8.0713,
-17.6153, -21.7492  and -22.4527 MeV respectively. The energy $E_{n \ell j}$(209) ($E_{n
\ell j}$(207)) represents the excitation energy centroid in $^{209}Pb$ ($^{207}Pb$)
around which the strength of a single-particle (single-hole) neutron orbital
(characterized by the quantum numbers $n$, $\ell$, and $j$) is fragmented.

\subsection{Results}

Table~\ref{tab:STAB1} depicts the theoretical binding energies, ${^{s.p.}B}_{th, n \ell
j}$, determined with the GAMOW code \cite{Gamow} for the single-particle neutron orbitals
$2g_{9/2}$, $1i_{11/2}$,  $3d_{5/2}$, $4s_{1/2}$, $2g_{7/2}$ and $3d_{3/2}$, using the
potential given by Eq. (1). Although the lead isotopes have been extensively investigated
since the beginning of nuclear physics, little experimental data on the fragmentation of
these orbitals is available in the literature  \cite{Martin}. These orbitals have been
predicted by the shell model and observed in single-particle stripping reactions on
$^{208}Pb$ \cite{Martin}. Based on a meticulous investigation of the
$^{208}Pb(d,p)^{209}Pb$ reaction at  $E_{lab} = 20$ MeV carried out by Kovar et al.
\cite{Kovar} and the results of different neutron stripping reaction studies compiled by
Martin \cite{Martin}, the strength of each one of these single-particle states may be
considered as mostly concentrated in one  excited state of the $^{209}Pb$ nucleus. In
order to determine each experimental binding energy ${^{s.p.}B}_{n \ell j}$ shown in
Table~\ref{tab:STAB1}, we assumed that each excitation centroid $E_{n \ell j}$(209) used
to evaluate Eq. (8), was equal to a particular $^{209}Pb$ excitation energy.

Table~\ref{tab:STAB2} contains different values determined  for the experimental binding
energy of the $1j_{15/2}$ orbital assuming three fragmentation schemes. From the study of
the $^{208}Pb(\alpha,{}^3He)$ reaction at $E_{lab}(\alpha)=183$ MeV, Massolo et al.
\cite{Massolo,Gales} demonstrated the strong population of the $1j_{15/2}$
single-particle level at 1.424 MeV excitation energy in $^{209}Pb$ and the existence of
fragments up to 5 MeV. In the first fragmentation scheme considered in
Table~\ref{tab:STAB2}, the $1j_{15/2}$ strength  is assumed to be concentrated completely
at 1.424 MeV excitation energy.  The schemes 2 and 3 include three fragments, in addition
to the 1.424 MeV state, that, according to Kovar et al. \cite{Kovar}, exhaust most of the
available $1j_{15/2}$ single-particle strength. Although these schemes include the same
fragments, their corresponding spectroscopic factors, extracted from a DWBA analysis of
the $^{208}Pb(d,p)$ reaction at $E_{lab}(\alpha)=20$ MeV, are different because the
deuteron break up potential was only included in the scheme 3.

Table~\ref{tab:STAB3} displays experimental and theoretical binding energies of the
single-hole neutron states in $^{207}Pb$ studied in the current work. Angular
distributions of cross sections and analyzing powers corresponding to the ground
($1/2^{-}$), 0.570 MeV ($5/2^{-}$) and 0.900 MeV ($3/2^{-}$) states in  $^{207}Pb$ were
measured by Matoba et al. \cite{Matoba} in a study  of the $^{208}Pb(p,d)^{207}Pb$
reaction at $E_{lab} = 65$ MeV. The spectroscopic factors obtained for these states
indicate that the strengths of the $3p_{1/2}$, $2f_{5/2}$ and $3p_{3/2}$ orbitals are
about 90 $\%$. These spectroscopic factors are consistent with the predictions of the
particle-vibration coupling model published by Majumdar \cite{Majumdar}. The experimental
fragmentation data corresponding to the $1i_{13/2}$, $2f_{7/2}$, $1h_{9/2}$ and
$1h_{11/2}$ orbitals were extracted from a study of the $^{208}Pb(^{3}He,\alpha)^{207}Pb$
reaction at $E_{lab} = 70$ MeV carried out by  Gal\`{e}s et al.~\cite{Gales}.

The theoretical binding energies presented in Tables \ref{tab:STAB1} and \ref{tab:STAB3}
for the $4s_{1/2}$, $3d_{3/2}$, $3d_{5/2}$, $3p_{1/2}$ and $3p_{3/2}$ orbitals are
consistent with those calculated by Wang et al.~\cite{Wang} using an elaborately
constructed multi-step potential to approximate the WS  +  SO potential specified in Eq.
(3). However, no agreement exists between any of the theoretical predictions and the
corresponding experimental value, similarly to what has been reported in the
literature~\cite{Blomqvist,Rost1,Dudek1,Dudek2,Volya}. As shown in Tables~\ref{tab:STAB1}
and \ref{tab:STAB2}, the theoretical binding energies of the single-particle neutron
states are in general less negative than the corresponding experimental values. The
opposite occurs for the single-hole neutron states displayed in Table~\ref{tab:STAB3}.
This is a consequence of how Schwierz et al. \cite{Volya} obtained the parameter values
$V_{0}= 52.06$ MeV and $\kappa = 0.639$, used in the current work to evaluate Eq. (7).
These parameters were optimized by using a set of $V_{0}$ and $\kappa$ values
corresponding to single-particle and single-hole neutron and proton states present in the
vicinity of the doubly-magic nuclei $^{16}O$, $^{40}Ca$, $^{48}Ca$, $^{56}Ni$,
$^{100}Sn$, $^{132}Sn$ and $^{208}Pb$.

\section{The Generalized Woods-Saxon plus Spin-Orbit Potential}
\subsection{Formalism}
In order to correct the differences mentioned in Section 2.2, between experimental and
theoretical energy values, we have replaced the traditional Woods-Saxon potential
$V_{WS}(r)$ contained in Eq. (1) by the Generalized Woods-Saxon potential
$V_{GWS}(r)$  \cite{Hamzavi,Bayrak1,Candemir,Berkdemir1,Berkdemir2,Bayrak2}, given by
\begin{equation}
V_{GWS}(r)= V_{WS}(r) + V_{SU}(r),
\end{equation}
where $V_{SU}(r)$ is a surface potential, proportional to a derivative of a Woods-Saxon
function, and given by
\begin{equation} \textstyle V_{SU}(r) = - \  C_{\ell j}
\frac{\exp\left(\frac{r\;-\;R}{a}\right)}{{(1+\exp\left(\frac{r\;-\;R}{a}\right))}^{2}} \
\ .
\end{equation}
The parameter $a$  was defined previously, $R$ is given by Eq. (6)  and the value of $C_{\ell
j}$ is calculated for each orbital when the theoretical binding energy coincides with the
experimental value.

\subsection{Results }
Table \ref{tab:STAB4} contains the $C_{\ell j}$ values obtained for all the
single-particle and single-hole neutron states studied here. The positive (negative)
signs of $C_{\ell j}$ determined for single-particle (single-hole) states are, in part, a
consequence of the parameterization Eq. (7),  and the associated parameter values
\cite{Volya}. We have also included in the table, the average $\overline{C}_{\ell}$ and
the difference $\Delta C_{\ell}$ of every pair of $C_{\ell j}$ values ($j=\ell\pm1/2$)
corresponding to the same quantum number $\ell$. They are given by

\begin{equation}
\textstyle \Delta C_{\ell} = C_{(j= \ell - 1/2)} - C_{(j= \ell + 1/2)} \ \ ,
\end{equation}
and
\begin{equation}
\textstyle \overline{C_{\ell}} = [C_{(j = \ell - 1/2)} + C_{(j = \ell + 1/2)}] / 2 \ \ .
\end{equation}

\begin{figure}[t]
\centering
\includegraphics[width=7cm]{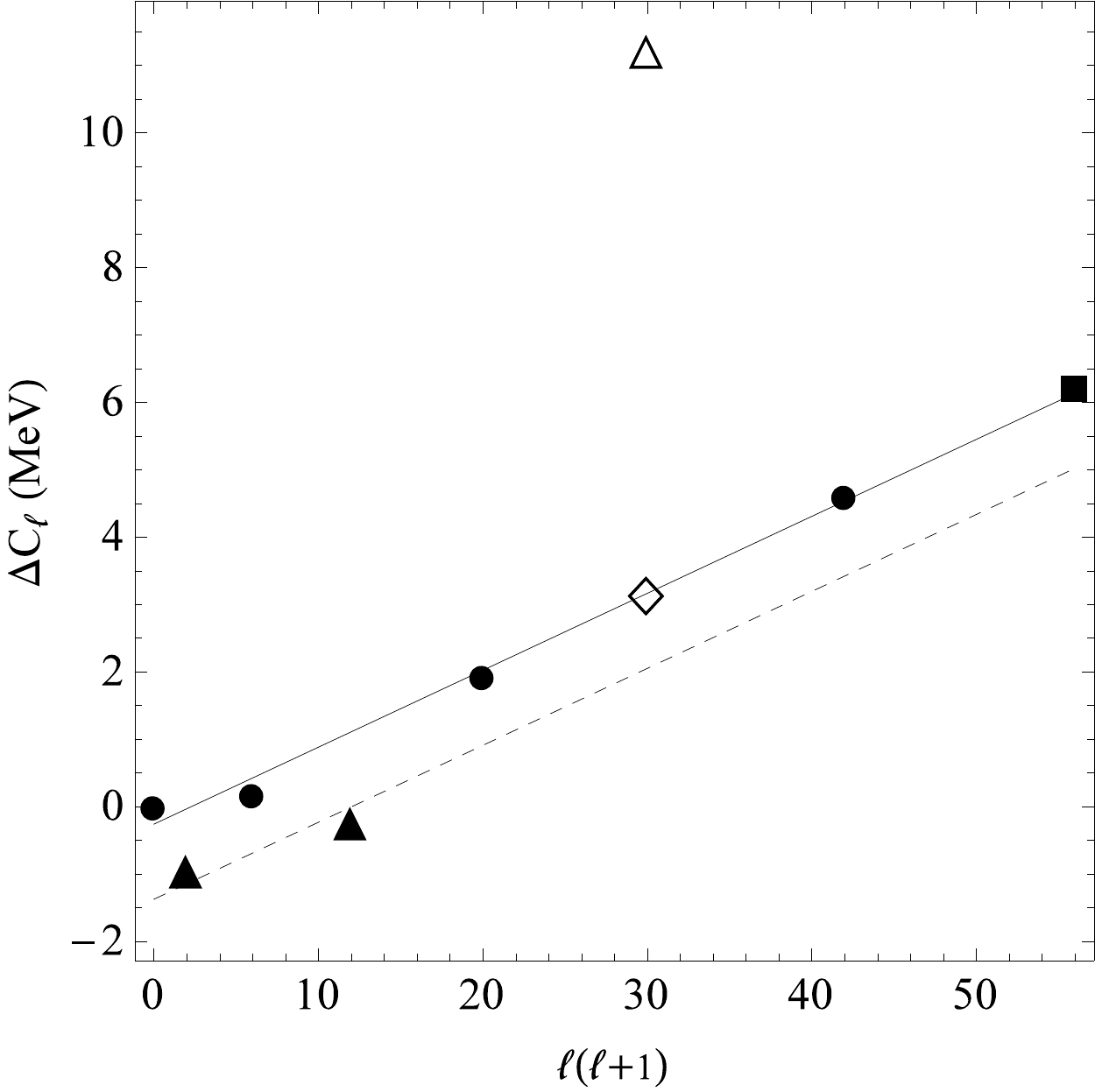}
\caption{$\Delta C_{\ell}$ = $C_{(j= \ell - 1/2)}$ - $C_{(j= \ell + 1/2)}$  vs.
$\ell(\ell+1)$. The solid line (Eq. (14)) was generated with the $\Delta C_{\ell}$ values
of the s, d, g and i orbitals (circles). The $\Delta C_{\ell}$ values obtained for the p
and f orbitals (solid triangles) lie close to a dashed line which is equal to the solid
line minus 1.1142 MeV. The determination of the empty triangle was based on the
excitation energy centroids measured by Gal\`{e}s \cite{Gales} for the h orbitals.  The
square and the diamond are explained in the text.} \label{Figure 1}
\end{figure}

\begin{figure}
\centering
\includegraphics[width=7cm]{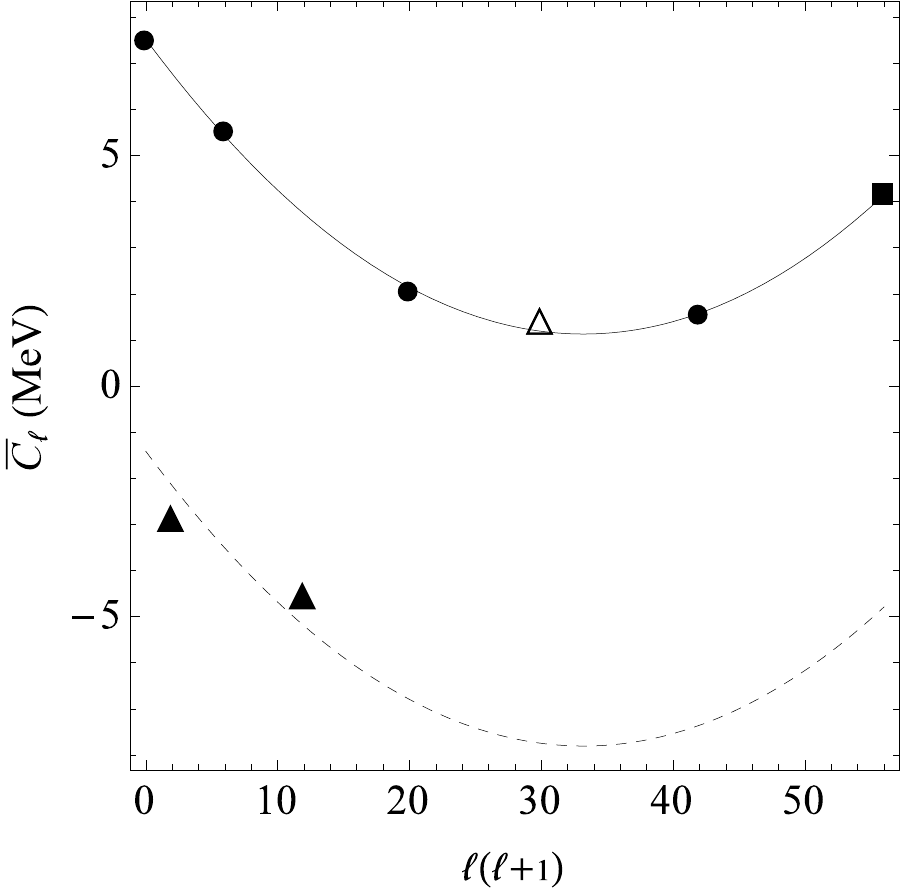}
\caption{$\overline{C}_{\ell}=(C_{j=\ell-1/2}+C_{j=\ell+1/2})/2$
 vs.
$\ell(\ell+1)$. The solid line (Eq. (15)) was generated with the $\overline{C}_{\ell}$
values of the s, d, g and i orbitals (circles). The $\overline{C}_{\ell}$ values obtained
for the p and f orbitals (solid triangles) lie close to a dashed curve which is equal to
the solid line minus 8.9355 MeV. The determination of the empty triangle was based on the
excitation energy centroids measured by Gal\`{e}s \cite{Gales} for the h orbitals.  The
square is explained in the text.} \label{Figure 2}
\end{figure}

Since the beginning of this investigation, we had the idea of finding a mathematical
expression to connect the strength $C_{\ell j}$ of the SU potential with a physical
quantity relevant to the problem. After a time consuming process of trial and error, we
finally realized that when the differences $\Delta C_{\ell}$ (see Table \ref{tab:STAB4})
associated to $\ell$ = 0, 2, 4 and 6 (solid circles in Fig.\ref{Figure 1}) are plotted as
a function of the emblematic quantity $\ell$($\ell$ + 1), they align along the straight
solid line shown in Fig. 1. Surprisingly, the four solid circles displayed in Fig.
\ref{Figure 2} (corresponding to the $\overline{C}_{\ell}$ values for $\ell$ = 0, 2, 4
and 6, plotted as a function of $\ell$($\ell$ + 1)) belong to the same parabola (solid
curve shown in Fig.\ref{Figure 2}). These results together with the fact that
$\overline{C}_{\ell}$ and $\Delta C_{\ell}$ are independent variables, support the
existence of a formula that describes the strength $C_{\ell j}$  in terms of
$\ell$($\ell$ + 1). The circles shown in Figs. \ref{Figure 1} and  \ref{Figure 2}, are
nicely described by the solid curves given by

\begin{equation}
\textstyle \Delta C_{\ell} = 0.1142 \ell (\ell + 1) - 0.2592 \ \ ,
\end{equation}
and
\begin{equation}
\textstyle \overline{C}_{\ell} = 0.0058 [\ell (\ell + 1)]^2 - 0.3851 \ell (\ell + 1) +
7.52  \ \ .
\end{equation}
The squares, triangles and diamonds included in the figures will be explained in detail
later.

The impressive agreement observed between the circles and the solid curves shown in Figs.
\ref{Figure 1} and  \ref{Figure 2}, allowed the determination of the $C_{\ell j}$ value
corresponding to the $1j_{13/2}$ orbital. This state lies above the shell  N = 126 - 184.
Evaluating Eqs. (14) and (15) for $\ell$ = 7, we obtained $\overline{C}_{(\ell= 7)}$ =
4.080 MeV and $\Delta C_{(\ell = 7)}$ = 6.127 MeV. In accordance with the notation used
in Eqs. (12) and (13), $C_{(j=13/2)}$ = $C_{(j= 7 - 1/2)}$ represents the $C_{\ell j}$
value of the $1j_{13/2}$ orbital. In principle, $C_{(j=15/2)}$ may be any of the three
$C_{\ell j}$ values displayed in Table IV for the $1j_{15/2}$ orbital. It is important to
realize that two $C_{(j=13/2)}$ values can be determined independently from Eqs. (12) and
(13) when one of the possible $C_{(j=15/2)}$ values is substituted into these equations.
This independence allowed us to rule out the values $C_{(j=15/2)}$ = -1.413 and -1.505
MeV (corresponding to the fragmentation schemes 2 and 3 respectively of the $1j_{15/2}$
orbital) since the two $C_{(j=13/2)}$ values obtained from these equations (for either
$C_{(j=15/2)}$ = -1.413 or -1.505 MeV)  were different. On the contrary, the
$C_{(j=13/2)}$ values obtained by evaluating Eqs. (12) and (13) with $C_{(j=15/2)}$ =
1.004 MeV (associated to the fragmentation scheme 1 of the $1j_{15/2}$ orbital), are
close (7.131 and 7.156 MeV). Substituting the average of these $C_{(j=13/2)}$ values
(7.143 MeV) and $C_{(j=15/2)}$ = 1.004 MeV into Eqs. (12) and (13), we obtained the values
$\overline{C}_{(\ell = 7)}$ = 4.074 MeV and $\Delta C_{(\ell =7)}$ = 6.139 MeV that are
close to the ones specified at the beginning of this paragraph. These new $\Delta
C_{(\ell =7)}$ and $\overline{C}_{(\ell = 7)}$ values are displayed as squares in Figs.
\ref{Figure 1} and \ref{Figure 2}. The discarding of negative $C_{(j=15/2)}$ values is
consistent with the general tendency observed in the $C_{\ell j}$ values corresponding to
single-particle neutron states of being positive.

The GAMOW code \cite{Gamow} was run for the $1j_{13/2}$ orbital, using the GWS+SO
potential with $C_{(j=13/2)}$ = 7.143 MeV . A positive binding energy of 4.473 MeV was
obtained for this unbound orbital which represents an excitation energy above the
$2g_{9/2}$ level of 8.410 MeV in $^{209}Pb$. In odd nuclei, such as the $^{209}Pb$
nucleus, some excited states can be described as admixtures of single-particle states or
quasiparticle states coupled to collective excitations of the even-even core \cite{Giai},
in this case the  $^{208}Pb$ nucleus. Previous works support the idea that the coupling
of a single-particle state with surface vibrations is mainly responsible for the damping
process of the single-particle mode \cite{Gales2,Giai2,Soloviev}. Based on the so called
quasiparticle-phonon model (QPM), Giai et al. \cite{Giai} found (Fig. 4 of Ref.
\cite{Giai}) that the strength of the $j_{13/2}$ state above 7 MeV excitation energy in
$^{209}Pb$ depends strongly on its coupling with the first $3^{-}$ and $5^{-}$
vibrational states in $^{208}Pb$. In particular, a bump obtained at 8.8 MeV excitation
energy was shown to be due to the coupling with the $5^{-}$ state. Similar calculations
were also carried out \cite{Giai} for the $k_{17/2}$ and $h_{11/2}$ states in $^{209}Pb$
and a  QPM inclusive spectrum that included the strength functions of the $k_{17/2}$,
$j_{13/2}$ and $h_{11/2}$ states was generated and compared to the experimental
$^{208}Pb(\alpha,{}^3He)^{209}Pb$ singles spectrum measured by Beaumel et
al.\cite{Beaumel} at 120 MeV incident energy and 0$^{\circ}$. The centroids of the
theoretical distribution of transfer cross sections between 6 and 12 MeV excitation
energy in $^{209}Pb$,  were shown to be shifted to higher energies by approximately 0.5
MeV in comparison with the centroids observed in the experimental distribution.   This
indicates that the excitation energy value predicted by us at 8.410 MeV for the
$1j_{13/2}$ state may be related to the theoretical $^{209}Pb$ fragment obtained by Giai
et al. \cite{Giai} at 8.8 MeV or represents the average excitation energy of all the
fragments calculated between  6 and 12 MeV.

The calculation of the $\Delta C_{(\ell=5)}$ value corresponding to the open triangle
displayed in Fig. \ref{Figure 1} is based on the $^{207}Pb$ fragmentations (see Table
\ref{tab:STAB3}) measured by  Gal\`{e}s et al. \cite{Gales} for the $1h_{9/2}$ and
$1h_{11/2}$ orbitals. This data point is too far from the solid curve. On the contrary,
the $\overline{C}_{(\ell=5)}$ value, shown also as an open triangle in Fig.~\ref{Figure
2}, is consistent with the $\overline{C}_{\ell}$ vs. $\ell (\ell + 1)$ value predicted by
the solid curve. This agreement facilitated the calculation of new $C_{\ell j}$ values
for the h orbitals that are different from those displayed in Table \ref{tab:STAB4}. For
this purpose, we substituted into Eqs. (12) and (13), the $\overline{C}_{(\ell=5)}$ value
corresponding to the open triangle of Fig. \ref{Figure 2} and the $\Delta C_{(\ell=5)}$
value predicted by the solid curve displayed in Fig. \ref{Figure 1} (open diamond). We
then run the GAMOW code \cite{Gamow} with the new $C_{\ell j}$ values to determine the
corresponding binding and excitation energies that are shown in Table~\ref{tab:STAB5}.
Although the excitation energy centroids $E_{n \ell j}$ displayed in this table for the
$1h_{9/2}$ and $1h_{11/2}$ orbitals (3.590 and 8.543 MeV respectively) differ from those
contained in Table ~\ref{tab:STAB3} (4.014 and 8.077 MeV respectively), the new centroids
lie inside the experimental excitation energy regions where Gal\`{e}s et al. \cite{Gales}
measured the fragments associated to the $1h_{9/2}$ ($E_{x}$ = 3.415 to 5.620 MeV) and
$1h_{11/2}$ ($E_{x}$ = 7.010 to 9.220 MeV) orbitals. The very strong background present
in the spectrum region where the fragments of these two h orbitals were detected \cite
{Gales}, suggests the convenience of carrying out new strength measurements to clarify
the disagreements mentioned here.

\begin{table}[t]
  \caption{  \label{tab:STAB4}$C_{\ell j}$ values of the surface potential $V_{SU}$ (Eq. (11)) that make coincide the theoretical and measured binding energies of the shell model states analyzed in the current work. $\overline{C}_{\ell}$ and $\Delta C_{\ell}$ are the average and the difference respectively of any pair of $C_{\ell j}$ values corresponding to a particular quantum number $\ell$.  The results labeled by a, b and c are the $C_{\ell j}$ values obtained for the fragmentation schemes 1, 2 and 3 respectively shown in Table~\ref{tab:STAB2} for the  $1j_{15/2}$ orbital.}
{\begin{tabular}{@{}*{5}{l}}
        \toprule
    orbital (n$\ell$j) & $\ell$ & $C_{\ell j}$ (MeV) & $\overline{C}_{\ell}$ (MeV)& $\Delta C_{\ell}$ (MeV)\\
        \colrule
        $4s_{1/2}$ & 0 & 7.509 & 7.509 & 0\\

        $3p_{1/2}$ & 1 & -3.307 & -2.842 & -0.930      \\
        $3p_{3/2}$ &   & -2.377 &       &      \\

        $3d_{3/2}$ & 2 & 5.626 & 5.536 & 0.180\\
        $3d_{5/2}$ &   & 5.446 &       &      \\

        $2f_{5/2}$ & 3 & -4.631 &-4.522& -0.218\\
        $2f_{7/2}$ &   & -4.413 &       &      \\

        $2g_{7/2}$ & 4 & 3.031 & 2.063 & 1.934\\
        $2g_{9/2}$ &   & 1.096 &       &      \\

        $1h_{9/2}$ & 5 & 7.055  &  1.436 &  11.238\\
        $1h_{11/2}$&   & -4.183 &       &      \\

        $1i_{11/2}$ & 6 & 3.866  & 1.562  & 4.608\\
        $1i_{13/2}$ &   & -0.742 &        &      \\

        $1j_{15/2}$ & 7 & 1.004$^a$  &       &      \\
                    &   & -1.413$^b$ &       &      \\
                    &   & -1.505$^c$ &       &      \\
        \botrule
        \end{tabular}}
      \end{table}

Using Eqs. (12) to (15), the strength $C_{\ell j}$ of an orbital characterized by j and
$\ell$ can be written as

\begin{equation} \label{eq:APP1.15}
\textstyle C_{\ell j} = C_{(j=\ell\pm1/2)} = \alpha_{(\pm)} [\ell (\ell + 1)]^2 +
\beta_{(\pm)} \ell (\ell + 1) + \gamma_{(\pm)} \ \ ,
\end{equation}
where $\alpha_{(\pm)}$ = 0.0058 MeV, $\beta_{(+)}$ = -0.4422 MeV, $\beta_{(-)}$ = -0.328
MeV, $\gamma_{(+)}$ = 7.6496 MeV and $\gamma_{(-)}$ = 7.3904 MeV. The convenience of
expression (16) is to have an easy way of determining the  $C_{\ell j}$ value required in
Eq. (11) to obtain the correct binding energy. Eq. (16) applies to all the orbitals
studied here except the p and f orbitals shown as solid triangles in Figs. \ref{Figure 1}
and \ref{Figure 2}. It would be advisable to revisit the investigation of these two
orbitals in the future: on the one hand, the experimental publication \cite{Matoba} from
which the data associated to these orbitals was extracted, seems to be solid and well
supported \cite{Majumdar}. On the other hand, the $\Delta C_{\ell}$  vs. $\ell (\ell +1)$
and $\bar{C}_{\ell}$ vs. $\ell (\ell +1)$  curves depicted as solid lines in Figs.
\ref{Figure 1} and \ref{Figure 2}, impressively reproduce the data corresponding to
$\ell$ =  0, 2, 4  and 6, and make reasonable predictions for $\ell$ = 5 and 7. We note,
without currently having a good explanation for it, that the p and f triangles lie close
to dashed curves obtained only by shifting the solid lines shown in Figs. \ref{Figure 1}
and \ref{Figure 2} by 1.1142 and 8.9355 MeV respectively. Based on this, the strength
$C'_{\ell j}$ of the p and d orbitals ($\ell$ = 1, 3) can be expressed in terms of
$C_{\ell j}$ as

\begin{equation} \label{eq:APP1.16}
\textstyle C'_{\ell j} = C'_{(j=\ell\pm1/2)} = C_{(j=\ell\pm1/2)} - \delta_{(\pm)} \ \ ,
\end{equation}
where $\delta_{+}$ = 8.3784 MeV and  $\delta_{-}$ = 9.4926 MeV.

\begin{table}[t]
\caption{\label{tab:STAB5}Binding energies ($B_{n \ell j}$) of $\ell$ = 5, 7 orbitals and
corresponding
         excitation energy centroids ($E_{n \ell j}$) in  $^{207}Pb$ and  $^{209}Pb$ respectively.
         The calculation of the $C_{\ell j}$ values is explained in the text.}
{\begin{tabular}{@{}*{5}{l}}
\toprule
        $\ell$ & orbital (n $\ell$ j)& $C_{\ell j}$ (MeV) & $B_{n \ell j}$ (MeV) & $E_{n \ell j}$ (MeV)\\
     \colrule
        7 & $1j_{15/2}$ & 1.004 & $-2.513$ & $1.424$ \\
          & $1j_{13/2}$ & 7.143 & $4.473$ & $8.410$ \\

        5 & $1h_{11/2}$ & 0.046 & $-15.911$ & $8.543$ \\
          & $1h_{9/2}$  & 2.826 & $-10.958$ & $3.590$ \\
\botrule
        \end{tabular}}
      \end{table}

\section{SUMMARY}
\indent We have used a generalized Woods-Saxon + a spin-orbit potential to reproduce the
experimental binding energies of single-particle and single-hole neutron orbitals
existing in the region around the doubly-magic $^{208}Pb$ nucleus. These orbitals are
contained in neutron shells that extend from $N = 82$ to $126$ (single-hole) and $N =
126$ to $184$ (single-particle).  A remarkable agreement has been found between the
binding energy values obtained by us for the $4s_{1/2}$, $3d_{3/2}$, $3d_{5/2}$,
$3p_{1/2}$ and $3p_{3/2}$ orbitals using the WS + SO part of the GWS + SO potential only, and
those determined in a previous investigation by Wang et al. \cite{Wang} following a novel
theoretical approach to solve the Schr\"{o}dinger equation. The WS and SO
parameterization and associated parameter values were taken from a recently published
work \cite{Volya}.\\
\indent Similarly to a common outcome reported in the literature
\cite{Blomqvist,Rost1,Dudek1,Dudek2,Volya,Wang}, the WS + SO part of the GWS + SO potential
did not produce by itself a global agreement between theoretical and experimental energy
values. We have accomplished this task with the addition of the SU potential. For each
orbital characterized by $\ell$ and j, we determined the strength $C_{(j=\ell\pm1/2)}$ of
this potential that makes the theoretical energy (calculated with the computer program
GAMOW \cite{Gamow}) coincide with the experimental value (determined from neutron
transfer
reaction experiments carried out on $^{208}Pb$ \cite{Martin,Kovar,Gales,Matoba}).\\
\indent A challenging result found in this investigation is the quadratic dependence of
the SU potential strength $C_{(j=\ell\pm1/2)}$ on the quantity $\ell(\ell+1)$. The
physical origin of this dependence needs to be understood. It is puzzling that while the
$C_{(j=\ell\pm1/2)}$ values corresponding to the s, d, g, h, i and j orbitals are nicely
described by the quadratic expression $C_{(j=\ell\pm1/2)}=\alpha_{(\pm)}[\ell (\ell +
1)]^2 + \beta_{(\pm)} \ell (\ell + 1)+\gamma_{(\pm)}$, the strengths of the p and f
orbitals ($C'_{(j=\ell\pm1/2)}$) are relatively close to the predictions of the shifted
expression $C'_{(j=\ell\pm1/2)} = C_{(j=\ell\pm1/2)} - \delta_{(\pm)}$. The different
behavior exhibited by the p and f orbitals in comparison with the general tendency
followed by
most of the orbitals needs further investigation.\\
\indent It would be important to find out if the quadratic $\ell(\ell+1)$ dependence of
$C_{(j=\ell\pm1/2)}$ obtained for $^{208}Pb$, also applies to single-particle and
single-hole neutron orbitals existing around other doubly-magic nuclei such as $^{16}O$,
$^{40}Ca$, $^{48}Ca$, $^{56}Ni$, $^{100}Sn$, $^{132}Sn$ and, if this is true, determine
if a global quadratic function with coefficients parameterized in terms of the neutron
magic numbers, could be generated. It would also be worthwhile to perform similar
investigations related to single-particle and single-hole proton orbitals.\\
\indent The explicit dependence of the SU potential strength on $\ell (\ell +1)$ allowed
the prediction of the binding energies of orbitals located inside and outside the neutron
shells analyzed. The excitation energy obtained for the $1j_{13/2}$ orbital may be
related to the excitation energy of a $1j_{13/2}$ fragment or the average of various
fragments calculated by Giai et. \cite{Giai} with the quasiparticle-phonon model where
each excited state originates from the coupling of a single-particle mode with a
vibrational state of the core. Based on this, we encourage the search of a formal
connection, if there is any, between the $\ell (\ell +1)$ dependence of the SU potential
and the quasiparticle-phonon description or any process that may clarify the physics
behind the SU potential included in this investigation. It is clear that further work is
required to understand formally from a microscopic point of view the interaction
mechanism responsable for the SU potential.\\
\indent Although the excitation energy centroids obtained for the h orbitals studied here
differ from those reported previously \cite{Gales}, their locations are inside the
excitation energy region where the relevant $^{207}Pb$ fragments were detected. Based on
the strong background present in the spectrum where the fragments were observed
\cite{Gales}, the remeasurement of the fragment strengths is advisable.\\
\indent To the best of our knowledge, this may be the first time that the binding
energies of single-particle and single-hole neutron states around  $^{208}Pb$ are
succesfully reproduced. This has been possible by the addition of a surface potential
(quadratically dependent  on $\ell (\ell +1)$) to the traditional Woods-Saxon plus
spin-orbit potential. In a previous investigation explained in the book by Nilsson et
al.~\cite{Nilsson}, the addition of a potential term (linearly dependent on $\ell (\ell
+1)$) to the harmonic oscillator potential make possible the reproduction of the level
ordering of the finite square well potential. In the same work, an additional inclusion
of a potential proportional to the dot product $\vec{\ell}\cdot\vec{s}$ resulted in a
successful reproduction of the single-neutron and single-proton level ordering measured
from the spectra of nuclei near $^{208}Pb$. In spite of the achievement mentioned in Ref.
\cite{Nilsson}, the agreement between
calculated and experimental energies was not satisfactory.\\
\indent It is well known that the harmonic oscillator basis has been extensively used in
shell models and mean field theories for spherical and deformed nuclei \cite{Wang}. Due
to the asymptotic behavior, the use of this basis, especially in cases such as exotic
nuclei, is questionable. The determination of the wavefunctions of a more realistic
potential like it may be the one used in this research would be a valuable task in future
investigations.

\section{Acknowledgments}

We wish to acknowledge the help received from graduate student Sttiwuer D\'{i}az in the
preparation of this document.

\end{document}